\documentclass[aps,prl,reprint,twocolumn,superscriptaddress]{revtex4-1}

\usepackage{euscript}
\usepackage{amssymb}
\usepackage{amsfonts}
\usepackage{amsbsy}
\usepackage{epsfig}
\usepackage{amsthm}
\usepackage{amscd}
\usepackage{amstext}
\usepackage{verbatim}
\usepackage{amsmath}
\usepackage{cancel}
\usepackage{capt-of}
\usepackage{empheq}
\usepackage{subfigure}
\usepackage{xcolor}

\usepackage[pdftex]{hyperref}
\hypersetup{ 
colorlinks=true, 
linkcolor=black, 
citecolor=red, 
}

\def\ben{\begin{equation}}
\def\een{\end{equation}}

 \let\t=\tau

\let\pa=\partial
\def\be{\begin{equation}}
\def\ee{\end{equation}}
\def\beq{\begin{equation}}
\def\eeq{\end{equation}}
\def\ba{\begin{array}}
\def\ea{\end{array}}

\def\dalemb#1#2{{\vbox{\hrule height .#2pt
       \hbox{\vrule width.#2pt height#1pt \kern#1pt
               \vrule width.#2pt}
       \hrule height.#2pt}}}

\newcommand{\bea}{\begin{eqnarray}}
\newcommand{\eea}{\end{eqnarray}}

\def\vep{{\varepsilon}}

\makeatletter
\newcommand*\bigcdot{\mathpalette\bigcdot@{.5}}
\newcommand*\bigcdot@[2]{\mathbin{\vcenter{\hbox{\scalebox{#2}{$\m@th#1\bullet$}}}}}
\makeatother

\renewcommand{\eqref}[1]{(\ref{#1})}

\begin{document}
\frenchspacing

\title{An upper bound on transport}

\author{Thomas Hartman}
\affiliation{Department of Physics, Cornell University, \\ Ithaca, New York, USA}
\author{Sean A. Hartnoll}
\affiliation{Department of Physics, Stanford University, \\
Stanford, California, USA}
\author{Raghu Mahajan}
\affiliation{Department of Physics, Stanford University, \\
Stanford, California, USA}

\begin{abstract}

The linear growth of operators in
local quantum systems leads to an effective lightcone even if the system is non-relativistic. We show that consistency of diffusive transport
with this lightcone places an upper bound on the diffusivity: $D \lesssim v^2 \t_\text{eq}$. The operator growth velocity $v$ 
defines the lightcone and $\t_\text{eq}$ is the local equilibration timescale, beyond which the dynamics of conserved densities is diffusive. We verify that the bound is obeyed in various weakly and strongly interacting theories. In holographic models this bound establishes a relation between the hydrodynamic and leading non-hydrodynamic quasinormal modes of planar black holes. Our bound relates transport data --- including the electrical resistivity and the shear viscosity --- to the local equilibration time, even in the absence of a quasiparticle description. In this way, the bound sheds light on the observed $T$-linear resistivity of many unconventional metals, the shear viscosity of the quark-gluon plasma and the spin transport of unitary fermions. 

\end{abstract}

\maketitle

\noindent {\it Operator growth, diffusion and local equilibration.---}In a local quantum spin system, operators can spread at most linearly in time, a fact that can be deduced via repeated commutation with the Hamiltonian \cite{Lieb1972, 2010arXiv1008.5137H}. This microscopic Lieb-Robinson bound establishes an effective `lightcone' for the propagation of signals, the spread of entanglement, and the generation of correlations, even in non-relativistic systems \cite{PhysRevLett.97.050401, eisert2015quantum}.  Although Lieb-Robinson theorems have been rigorously proven only in certain classes of models, there is a large and growing body of evidence that local quantum systems obey a finite speed limit more generally.  The linear spatial growth of entanglement and correlation with time has been widely observed in analytic \cite{Calabrese:2005in,Calabrese:2006rx}, numerical \cite{hub, PhysRevB.79.155104, PhysRevB.85.085129, PhysRevLett.111.127205, Bohrdt:2016vhv, Leviatan:2017vur} and experimental \cite{cheneau2012light, langen2013local, jurcevic2014quasiparticle} modelling of one dimensional systems, as well as in higher dimensional models that can be studied through holographic duality \cite{Albash:2010mv,Hartman:2013qma,Liu:2013iza,Liu:2013qca}. We will denote the velocity defining the operator growth lightcone, that bounds any spread of entanglement or correlation, by $v$.

Another property of (ergodic) local systems is that conserved densities diffuse at late times and long distances, see e.g. \cite{KADANOFF1963419}. 
Diffusion implies that the retarded Green's function for the conserved density $n$ takes the late time and long wavelength form
\be\label{eq:nn}
\langle [n(t,x),n(0,0)] \rangle \propto \nabla^2 \frac{e^{-x^2/(4 D t)}}{t^{d/2}} \,,
\ee
for $t \gtrsim \tau_\text{eq}$ and $|x| \gtrsim \ell_\text{eq}$.
Here $D$ is the diffusion constant and $d$ is the dimensionality of space. We also introduced the local equilibration time $\tau_\text{eq}$ and the local equilibration lengthscale $\ell_\text{eq}$. These are the scales beyond which the hydrodynamic derivative expansion holds, and will play a central role in our discussion.

We will combine the above two facts to derive an upper bound on $D$. In the context of relativistic theories, a large literature exists on the tension between diffusion and the usual relativistic lightcone, see \cite{Baier:2007ix,Romatschke:2009im} for discussions. We elaborate on this connection below. The logic of our argument is more closely related to well-known bounds on the low energy coupling constants of relativistic theories that follow from causality and unitarity  \cite{Adams:2006sv,Camanho:2014apa,Hartman:2015lfa}.  
In some cases, such as \cite{Camanho:2014apa,Afkhami-Jeddi:2016ntf}, the bounds arise because a coupling in the low energy effective theory violates causality, which is allowed only if the coupling is small enough to push the violation beyond the cutoff scale. The existence of an operator growth lightcone makes it possible to apply the same logic to non-relativistic theories and leads to the bound on $D$.

Diffusion constants directly control transport via Einstein relations \cite{KADANOFF1963419}. For example, the electrical conductivity is related to charge diffusivity by $\sigma = \chi \, D$, where the charge compressibility $\chi = \pa n/\pa \mu$. We will also apply the bound to spin, heat and momentum transport.

\begin{figure}[h]
\centering
\includegraphics[width=3.5in]{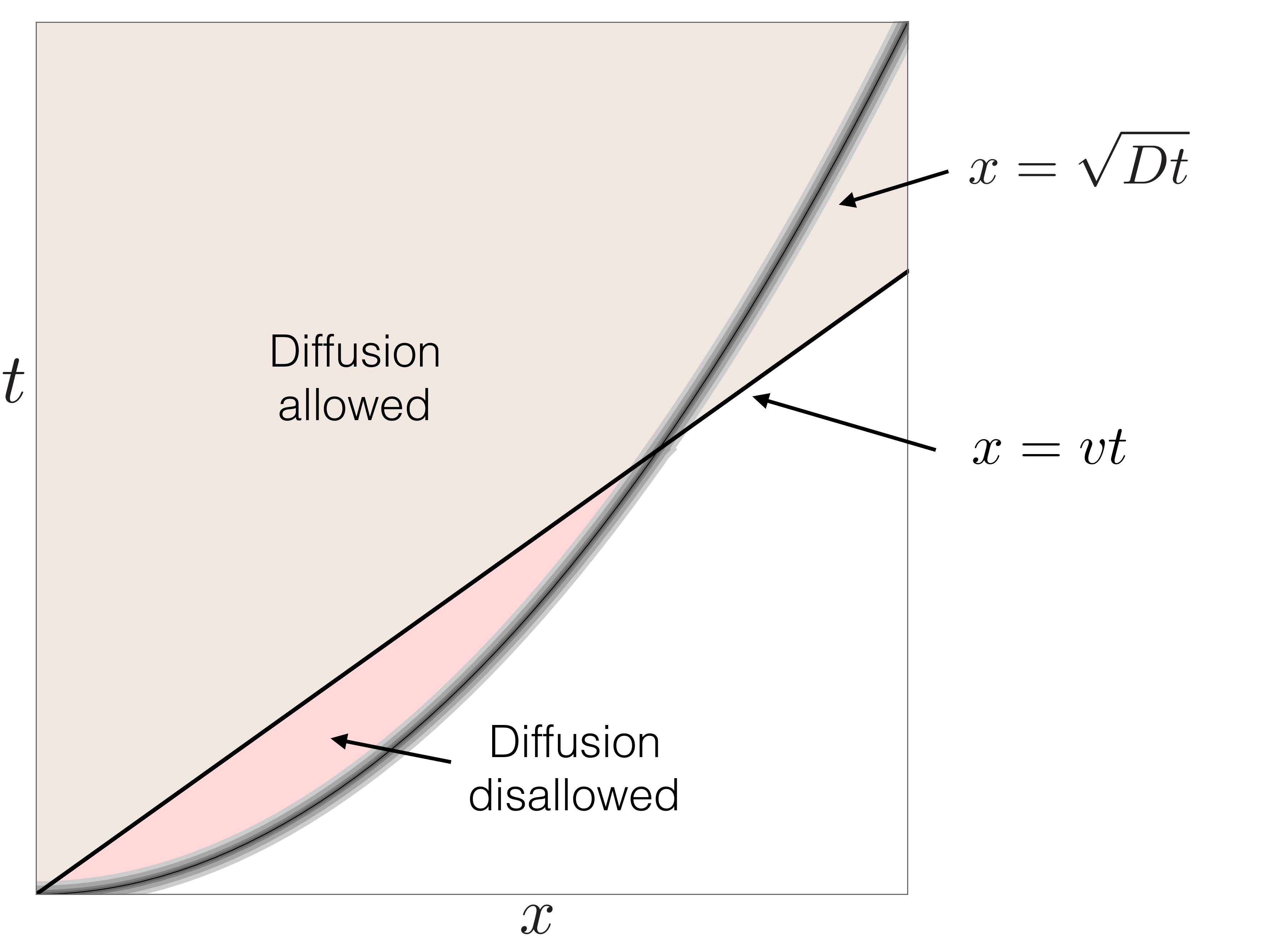}
\caption{The diffusive Green's function is large outside the operator growth lightcone at early times. It follows that the true Green's function must not be diffusive over that regime.
}
\label{fig:intersect}
\end{figure}

{\it Derivation of the bound.---}The diffusive Green's function (\ref{eq:nn}) is large for $|x| \lesssim \sqrt{D t}$. The numerical prefactor in this inequality will not be fixed by our argument. At early times, the region of spacetime $|x| \lesssim \sqrt{D t}$ includes points that are at $|x| > v t$, outside of the operator growth lightcone. See figure \ref{fig:intersect}.
Thus, it seems that diffusion allows us to send signals outside of the lightcone.
However, the Green's function only becomes diffusive at times
$t \gtrsim \t_\text{eq}$. The tension between diffusive behavior and the linear-in-time growth of operators is therefore avoided if
\be\label{eq:Dbound}
D \lesssim v^2 \t_\text{eq} \,.
\ee
The simple observation (\ref{eq:Dbound}) is our main result, relating three independently defined quantities.
Indeed, $D, \tau_\text{eq}$ and $v$ have been independently computed in several models \cite{PhysRevLett.111.127205, Bohrdt:2016vhv, Leviatan:2017vur}, and the bound (\ref{eq:Dbound}) is found to hold. The numerical factor in \eqref{eq:Dbound} is undetermined because the edge of the diffusive region is defined in the low energy, long wavelength theory, so it is smeared by an amount $\tau_\text{eq}$ in the time direction and $\ell_\text{eq}$ in the space direction. 
The argument leading to (\ref{eq:Dbound}) assumes that the local equilibration lengthscale $\ell_\text{eq} \sim v \tau_\text{eq}$
and that the operator growth lightcone is enforced even at early times $t \sim \tau_\text{eq}$,
as is indeed observed \cite{hub, PhysRevB.79.155104, PhysRevB.85.085129, cheneau2012light}. 

In many-body localized systems, the `lightcone' curve becomes $t = e^{x/\xi}$ \cite{PhysRevB.90.174202,2014arXiv1412.3073K}. Our argument then leads to the known fact that the diffusivity $D=0$.

In weakly interacting systems, {\it lower} bounds on the diffusivity follow from applying an uncertainty principle argument to the lifetime of quasiparticles \cite{Kovtun:2004de, Hartnoll:2014lpa}.
It remains possible that lower bounds on diffusivities also exist away from weak coupling, as has been conjectured in \cite{Kovtun:2004de, Hartnoll:2014lpa, Blake:2016wvh}.
The bound (\ref{eq:Dbound}) goes in the same direction as the upper bound on diffusivity found in \cite{Lucas:2016yfl, Gu:2017ohj} for certain models with long-wavelength inhomogeneities.

{\it Microscopic vs.~low energy velocities.---}The velocity $v$ that defines the operator growth lightcone is a microscopic velocity. On a lattice, it is set by the lattice spacing $a$ and a characteristic microscopic energy scale $J$ according to $v \sim J a/\hbar$ \cite{2010arXiv1008.5137H}. If the typical velocity of low energy carriers of the conserved density is substantially lower than this velocity, the bound (\ref{eq:Dbound}) is very weak.

There are two important cases where the low energy excitations that carry the conserved densities do in fact have a microscopic velocity. The first are relativistic systems where the characteristic velocity is the speed of light. The second are degenerate Fermi systems where 
the characteristic velocity is the Fermi velocity $v \sim v_F \sim J a/\hbar$. For these cases we can expect the diffusion bound (\ref{eq:Dbound}) to be a nontrivial constraint.

In other circumstances the velocity of low energy excitations will be set by the low temperature $T$, with e.g. $v_T \sim T^{1-1/z}$. This is much slower than any microscopic velocity when the dynamical critical exponent $z>1$.
It may be possible to strengthen the bound in such cases. A well-defined velocity is the butterfly velocity $v_B$ that controls the chaotic growth of the commutator $\langle -[A(t,x),A(0,0)]^2 \rangle \sim e^{(t - x/v_B)/\tau_L}$ \cite{Shenker:2013pqa, Roberts:2014isa}. Here $\tau_L$ is the Lyapunov timescale \cite{kitaevtalks,Maldacena:2015waa}. The butterfly velocity tracks the characteristic velocity of low energy excitations, with $v_B \sim v_T$ \cite{Blake:2016wvh, Roberts:2016wdl}, and indeed has the flavor of an effective state-dependent Lieb-Robinson or operator growth velocity \cite{Roberts:2014isa,Alishahiha:2014cwa,Fonda:2014ula,Roberts:2016wdl, Mezei:2016wfz, Qi:2017ttv}. Furthermore, computations in specific holographic \cite{Blake:2016wvh, Blake:2016sud,Blake:2016jnn, Baggioli:2016pia, Blake:2017qgd, Baggioli:2017ojd,Wu:2017exh}, perturbative quantum field theoretic \cite{Patel:2016wdy, Aleiner:2016eni, Chowdhury:2017jzb, Patel:2017vfp}, and 1+1 dimensional \cite{Gu:2016oyy, Davison:2016ngz, Bohrdt:2016vhv} systems have found diffusion controlled by the butterfly velocity, with $D \sim v_B^2 \tau_L$.

The timescales $\tau_L$ and $\tau_\text{eq}$ are defined in quite different ways. In many cases $\tau_\text{eq} \sim \tau_L$, but below we give an example with $\tau_\text{eq} \gg \tau_L$. The definition of $\tau_\text{eq}$ as the timescale beyond which conserved quantities diffuse was essential to make the general argument for (\ref{eq:Dbound}) above. Indeed there are known cases with $D \ll v_B^2 \tau_L$ \cite{Lucas:2016yfl, Gu:2017ohj}, as well as cases with $D \gg v_B^2 \tau_L$ \cite{Blake:2016wvh, Blake:2017qgd}.
The natural conjecture for a stronger bound extending (\ref{eq:Dbound}) is then $D \lesssim v_B^2 \tau_\text{eq}$ \footnote{There are holographic \cite{Blake:2016wvh, Blake:2017qgd} and non-holographic \cite{steve} large $N$ models with a `fast' diffusion mode, seemingly with $D \gg v_B^2 \tau_\text{eq}$, as well as models with sound modes that move faster than the large $N$ butterfly velocity. In these cases the existing computations of the butterfly velocity may get significant corrections at subleading order in $1/N$. We thank Steve Shenker and Douglas Stanford for discussions of this point.}.

{\it Quasiparticle systems and a bound on $\tau_\text{eq}$.---}If a system has long lived quasiparticles with velocity $v_\text{qp}$ and lifetime $\tau_\text{qp}$, the diffusivity is given by $D \sim v_\text{qp}^2 \tau_\text{qp}$. Equilibration is slow in such systems because interactions are weak and typically $\tau_\text{eq} \sim \tau_\text{qp}$. Our bound (\ref{eq:Dbound}) then simply becomes $v_\text{qp} \lesssim v$,
which certainly has to be true, else the quasiparticles could carry signals outside the operator growth lightcone. 

In a weakly coupled system the diffusivity can be lower bounded by the uncertainty principle applied to the single-particle excitations \cite{Kovtun:2004de, Hartnoll:2014lpa}. Define the effective mass of the quasiparticles by $\vep_\text{qp} = m_* v_\text{qp}^2$, where $\vep_\text{qp}$ is the quasiparticle energy. 
Then we can write
\be\label{eq:lower}
D \sim v_\text{qp}^2 \tau_\text{qp} \sim \frac{\vep_\text{qp} \tau_\text{qp}}{m_*} \gtrsim \frac{\hbar}{m_*} \,.
\ee
Such quantum-limited diffusion has been directly observed in ultracold atomic Fermi liquids. Approximate saturation of (\ref{eq:lower}) occurs as unitarity is approached and the quasiparticle description breaks down \cite{Sommer2011, Bardon722, PhysRevLett.114.015301, PhysRevLett.118.130405}. Compatibility of the lower bound (\ref{eq:lower}) with the upper bound (\ref{eq:Dbound}) requires that in a quasiparticle system
\be\label{eq:taueq}
\tau_\text{eq} \gtrsim \frac{\hbar}{m_* v^2} \,.
\ee
At low temperatures, the bound (\ref{eq:taueq}) is weaker than the bound $\tau_\text{eq} \gtrsim \hbar/k_B T$ conjectured in \cite{subirbook}.

{\it Holographic theories and quasinormal modes.---}The best studied set of holographic theories are conformal field theories (CFTs) in $d$ spatial dimensions placed at a nonzero temperature. 
In a CFT, the transverse momentum diffusivity is given by the shear viscosity according to $D = \eta/(s T)$ \cite{Son:2007vk}. Throughout this subsection only we set $c= k_B = \hbar =1$. The bound (\ref{eq:Dbound}) can therefore be written
\be\label{eq:etas}
\frac{\eta}{s} \lesssim v^2  \, T \, \tau_\text{eq} \,.
\ee
In a nonzero temperature CFT, $v=1$ and $T \tau_\text{eq}$ is a temperature-independent number.

The local equilibration timescale $\tau_\text{eq}$ is the lifetime of the longest-lived non-hydrodynamic excitation. The damped excitations of black holes are called quasinormal modes, the singularities of the frequency-space retarded Green's function.
Thus we can write \cite{Horowitz:1999jd}
\be\label{eq:qnm}
\tau_\text{eq} = \frac{1}{\text{Im} \, \omega^\text{qnm}} \,,
\ee
where $\omega^\text{qnm}$ is the complex frequency of the non-hydrodynamic quasinormal mode closest to the real axis. In fact (\ref{eq:qnm}) is not quite enough. The quasinormal frequencies depend on the wavevector $k$ and at large $k$ the quasinormal modes become arbitrarily close to the real axis (see e.g. \cite{Festuccia:2008zx}). The bound (\ref{eq:Dbound}) is only concerned with the 
local thermalization timescale at long distances where hydrodynamics is valid. Therefore, we can minimize (\ref{eq:qnm}) over $k$ restricted to $k < \ell_\text{eq}^{-1} \sim (v \tau_\text{eq})^{-1}$. In practice, the lowest non-hydrodynamic quasinormal frequencies do not have a strong $k$ dependence for small $k$. Thus it is sufficient to put $k=0$ in (\ref{eq:qnm}).

{\it Einstein gravity.} The leading non-hydrodynamic quasinormal modes in the shear sector of Einstein gravity have been studied in detail in \cite{Morgan:2009pn}.
Using (\ref{eq:qnm}), together with the $k=0$ shear sector quasinormal frequencies computed in \cite{Morgan:2009pn}, we obtain $T \tau_\text{eq} \approx 0.09, 0.12, 0.15$ in $d=2,3,4$, respectively. These values are in agreement with the bound (\ref{eq:etas}), with $v=1$ and given the value of $\eta/s = 1/(4 \pi) \approx 0.08$ \cite{Policastro:2001yc, Kovtun:2004de}.

{\it Higher derivative gravity.} In higher derivative theories of gravity $\eta/s$ can become parametrically large. Two examples of such theories were studied in \cite{Grozdanov:2016vgg}, where it was shown that the ratio $\eta/(s T \tau_\text{eq})$ nonetheless converges to a constant such that (\ref{eq:etas}) holds with $v=1$. These theories are toy models for exploring the possible behaviors of retarded Green's functions. However, large higher derivative corrections typically lead to inconsistencies in the full theory  \cite{Brigante:2008gz,Hofman:2009ug,Camanho:2014apa}.

{\it Linear axion spacetimes.} In `linear axion' spacetimes a parameter $m$ controls the strength of momentum degradation due to broken translations \cite{Andrade:2013gsa}.
This determines the leading non-hydrodynamic quasinormal mode, which has been characterized in \cite{Davison:2014lua}. The heat diffusion constant and butterfly velocity have been obtained in \cite{Blake:2016sud}.
When $m$ is small $D/\tau_\text{eq} \sim v_B^2 \sim 1$. When $m$ is large $D/\tau_\text{eq} \sim v_B^2 \sim T/m$. 
The bound (\ref{eq:Dbound}) is found to be saturated in both limits if we use the butterfly velocity $v \to v_B$. If we use the microscopic lightcone velocity, $v = 1$, then the bound is always satisfied but is far from saturated with strong momentum relaxation. In this example replacing $\tau_\text{eq} \to \tau_L$, the Lyapunov time, in the bound \eqref{eq:Dbound} would not be valid. At weak momentum relaxation $\tau_L \ll \tau_\text{eq}$ \cite{Blake:2016sud}, and hence $D \gg v_B^2 \tau_L \sim v^2 \tau_L$ in that limit.

{\it The case of a long-lived momentum.---}With weakly broken translation invariance, the total momentum decays at a `transport rate' rate $\tau_\text{tr}^{-1}$ that is slow compared to all other non-hydrodynamic excitations, that instead decay at the rate $\tau_\text{eq}^{-1}$.
When $\tau_\text{tr}^{-1} \ll \tau_\text{eq}^{-1}$ the decay of the total momentum can itself be described within hydrodynamics.
One obtains (e.g. \cite{Davison:2014lua, KADANOFF1963419})\footnote{In \cite{Davison:2014lua} the terms in brackets in the denominator of (\ref{eq:relaxtime}) include an extra $k^2$ term that describes attenuation of sound in the large $\tau_\text{tr} \omega$ limit. That term is subleading in the regimes of interest to us and so we have dropped it for clarity. There can also be additional `incoherent' diffusive modes, that are independent of momentum relaxation \cite{Davison:2015taa}. These lead to an additive constant $\sigma_o$ in the conductivity  (\ref{eq:relaxtime}).}
\be\label{eq:relaxtime}
\sigma(\omega,k) = \frac{- i \omega \, G^R_{nn}(\omega,k)}{k^2} = \frac{i \omega \, \chi D}{i \omega (1 - i \tau_\text{tr} \, \omega) - D k^2} \,.
\ee
At $k=0$ this is the standard Drude formula for the conductivity. The behavior at nonzero $k$ depends upon $\tau_\text{tr}\, \omega$. At the lowest frequencies, $\tau_\text{tr} \, \omega \ll 1$, equation (\ref{eq:relaxtime}) describes diffusion of charge. At higher frequencies with $\tau_\text{tr} \, \omega \gg 1$ there is a linearly dispersing mode with $\omega^2 = (D/\tau_\text{tr}) \, k^2$. Because $\tau_\text{tr}$ is large, this latter regime is still within the validity of a hydrodynamic description. Signals cannot propagate faster than the operator growth velocity and hence 
\be
D \leq v^2 \tau_\text{tr} \,,
\ee
recovering the bound (\ref{eq:Dbound}) with $\tau_\text{eq} \to \tau_\text{tr}$, as $\tau_\text{tr}$ controls the crossover from diffusion to sound, and now with a precise numerical factor.

It has long been appreciated in the context of relativistic hydrodynamics that there is a tension between diffusion and causality \cite{Muller:1967zza,Israel:1976tn}. An overview of the issues can be found in \cite{Baier:2007ix,Romatschke:2009im}. Causality violation in the diffusion equation is  superficially resolved by including a new transport coefficient $\tau_\pi$ at second order in the hydrodynamic derivative expansion. This leads to the same Green's function as \eqref{eq:relaxtime}, with $\tau_\text{tr}$ replaced by $\tau_\pi$. Argumentation similar to that in the previous paragraph, with $v \to c$, the speed of light, has then been used to bound diffusion in that context, leading to the suggestion $D \leq c^2 \tau_\pi$. However, as emphasized in \cite{Baier:2007ix}, such arguments are generically uncontrolled as they involve keeping one out of infinitely many non-hydrodynamic modes. In the discussion above, in contrast, we have considered the case of a single long-lived non-hydrodynamic mode that can arise due to weakly broken translation invariance.

Similar considerations to the above apply to superfluids, with the supercurrent operator weakly relaxed by the transverse motion of vortices \cite{Davison:2016hno}.

{\it Resistivity of non-quasiparticle metals.---}The resistivity of many families of strongly correlated materials exhibits the temperature dependence $\rho \sim T$. Many of these are bad metals, with large resistivities that are incompatible with a quasiparticle description \cite{PhysRevLett.74.3253, RevModPhys.75.1085, MIR}. The conventional Drude formula $\rho =1/\sigma = m/(n e^2 \tau)$ is therefore not applicable. Nonetheless, the resistivity of these same materials does appear to be associated with an underlying timescale $1/\tau_\text{eq} \sim T$ extracted from, for example, the width of peaks in the optical conductivity $\sigma(\omega)$. See e.g.
\cite{Sachdev:2011cs, andy, Hartnoll:2014lpa, Delacretaz:2016ivq} for overviews of strongly correlated materials with $\rho \sim 1/\tau_\text{eq} \sim T$. The cuprates are especially well-characterized in this regard, with the $\tau_\text{eq} \sim \hbar/(k_B T)$ timescale widely seen in optical data \cite{PhysRevB.42.6342,PhysRevB.41.11237,OPTICAL1} as well as in single-particle observables \cite{ARPES1}.

The diffusivity bound (\ref{eq:Dbound}) implies a lower bound on the electrical resistivity:
\be\label{eq:rr}
\rho \gtrsim \frac{1}{\chi} \frac{1}{v^2} \frac{1}{\tau_\text{eq}} \,.
\ee
This expression is a non-quasiparticle generalization of the conventional Drude formula, relating the resistivity to a timescale --- as seen in the $T$-linear resistivity materials. It becomes the Drude formula in the quasiparticle limit.
At degenerate temperatures $k_B T \ll E_F$, with $E_F$ the Fermi energy, the important temperature dependence in the resistivity bound (\ref{eq:rr}) indeed comes from $\tau_\text{eq}$. The susceptibility $\chi \sim n e^2/E_F$ and the velocity $v \sim v_F$ are temperature-independent \footnote{A temperature dependence of $v$ can arise, even at degenerate temperatures, if an additional microscopic velocity, such as the sound speed, also contributes to the operator growth velocity \cite{Zhang08052017}. This may be related to kinks in $v_F$ revealed by photoemission \cite{Lanzara2001,Zhou2003}.}. $v_F$ can be extracted from angle-resolved photoemission data that reveals a well-defined single-particle peak in momentum space \cite{Valla2110}, despite broadening in frequencies \cite{PhysRevLett.86.4362}.

{\it Spin diffusion in unitary Fermi liquids.---}The relaxation to equilibrium of a spin imbalance in a 3d unitary Fermi liquid was characterized in \cite{Sommer2011}. In addition to spin diffusion at late times, the experiment also sees frictional damping at an intrinsic (geometry-independent) rate called the spin drag coefficient $\Gamma_\text{sd}$, giving the equilibration time $\tau_\text{eq} \sim 1/\Gamma_\text{sd}$. At low temperatures the spin diffusion constant is found to be $D_\text{s} \approx 6 \, \hbar/m$ and the equilibration time $\tau_\text{eq} \approx 10 \, \hbar/E_F$. In this degenerate regime we can estimate $v^2 \sim v_F^2 \sim E_F/m$. It follows that the diffusivity bound (\ref{eq:Dbound}) holds and is approximately saturated.
The values of $D_\text{s}$ and $1/\Gamma_\text{sd}$ just quoted are both overestimated by the same factor due to averaging over the inhomogeneous trapping potential. Allowing for this geometric effect, the homogeneous diffusivity is estimated in the supplementary material of \cite{Sommer2011} to be $D_\text{s} \sim \hbar/m$, see also \cite{PhysRevLett.107.255302, PhysRevLett.109.195303}, closer to the diffusivities discussed below.

Transverse spin diffusivity has also been measured in a 2d unitary Fermi liquid to be $D_0^\perp \approx 2 \hbar/m$ \cite{PhysRevLett.118.130405}.
That work did not obtain an independent relaxation rate. However, a separate experiment on a very similar system measured
the damping rate $\Gamma_Q$ of a quadrupole mode \cite{Vogt:2011np}. In the strongly interacting regime the intrinsic relaxation timescale is given by $\tau_\text{eq} \sim \Gamma_Q/\omega_\perp^2$, where $\omega_\perp$ is the harmonic trapping frequency \cite{PhysRevA.87.043612}. 
The low temperature, strongly interacting data in \cite{Vogt:2011np} then leads to $\tau_\text{eq} \approx 25 \, \hbar/E_F$. Again estimating $v^2 \sim v_F^2 \sim E_F/m$ at low temperatures, we see that the diffusivity bound (\ref{eq:Dbound}) is satisfied, and is not especially close to saturation in this case. This comparison may be imperfect as the equilibration timescale and diffusivity pertain to different modes.

{\it Viscosity.---}The conjectured lower bound on the shear viscosity \cite{Kovtun:2004de} has motivated measurements of the viscosity of strongly interacting media \cite{Cao58, SONG2013114c, Adams:2012th}.

In the quark-gluon plasma,
$\tau_\text{eq} \sim 0.6$ fm/c \cite{Heinz:2001xi}. Using $T \sim 360 $MeV and setting $v=c$, the momentum transport bound (\ref{eq:etas}) becomes $\eta/s \lesssim 1.1 \, \hbar/k_B$, which is consistent with the measurement $\eta/s \sim 0.15 \, \hbar/k_B$ \cite{Song:2011hk,SONG2013114c}.

For Galilean systems, such as ultracold atomic Fermi liquids, the momentum transport bound becomes $D = \eta/(n m) \lesssim v^2 \tau_\text{eq}$.
The low temperature viscosity of a Fermi liquid tuned to unitarity has been measured to be $\eta/n \approx 0.4 \, \hbar$ \cite{Cao58, Luo2009}. An independent measurement of the local equilibration time in the same elliptic flow 
experiments from which the shear viscosity is extracted is needed to verify our bound in these systems.

{\it Conclusion.---}We obtained the bound (\ref{eq:Dbound}) relating three independent quantities: diffusivity, equilibration timescale and lightcone velocity. Empirically, the bound holds in a wide variety of physical systems at strong and weak coupling. These results motivate a more systematic and simultaneous determination of the three quantities in e.g. ultracold atomic liquids and unconventional metals.

\bigskip

\begin{acknowledgments}

It is a pleasure to acknowledge a helpful early comment from Patrick Hayden, and insightful comments on an earlier draft of this paper from Blaise Gouteraux,  Andrew Lucas, Steve Shenker, Douglas Stanford, Joseph Thywissen and Xiaoliang Qi. The work of TH is supported by DOE grant DE-SC0014123. The work of SAH and RM is supported by a DOE Early Career Award.

\end{acknowledgments}

\end{document}